\documentclass[a4paper,11pt]{article}
\pdfoutput=1 % if your are submitting a pdflatex (i.e. if you have
\usepackage{jinstpub} 
\usepackage{lineno}
\usepackage{siunitx}
\usepackage{graphicx}
\usepackage{subcaption}

\title{\boldmath Development of Readout Electronics for a High-Speed Event-Driven Neutron Imaging Detector Based on Timepix4}

\author[a,b,c]{Q. Li}
\author[a,b,1]{H. Liu\note{Corresponding author.}}
\author[a,b,c]{D. Cai}
\author[b,d]{H. Guo}
\author[a,b]{X. Jiang}
\author[a,b]{H. Teng}
\author[a,b,c]{K. Wang}
\author[a,b]{X. Wang}
\author[a,b]{S. Wang}
\author[a,b]{Z. Sun}
\author[a,b]{Y. Zhao}
\author[a,b]{J. Zhou}

\affiliation[a]{Institute of High Energy Physics, Chinese Academy of Sciences, Beijing, 100049, China}
\affiliation[b]{Spallation Neutron Source Science Center, Dongguan, Guangdong, 523803, China}
\affiliation[c]{University of Chinese Academy of Sciences, Beijing 100049, China}
\affiliation[d]{University of Electronic Science and Technology of China, Chengdu, Sichuan 610054, China}
\emailAdd{hbliu@ihep.ac.cn}

\abstract{
As the Chinese Spallation Neutron Source enters Phase II, the increase in proton beam power will lead to a further boost in the intensity of pulsed neutron beams. To address the demand for higher event-rate readout electronics for energy-resolved neutron imaging detectors, we have developed a high-performance readout electronics system based on the Timepix4 chip. The prototype electronics system comprises a Timepix4 chip board and a high-performance digital board, which are interconnected through a custom FMC interface. The advantage of this system is its ability to achieve the full bandwidth readout of 160 Gbps for a single Timepix4 chip. The electronics system, based solely on a single ZYNQ-MPSOC chip, is capable of fully meeting the required performance specifications within a compact form factor of \(8 \text{ cm} \times 30 \text{ cm}\). Furthermore, the system features a high-capacity external SODIMM memory interface (supporting up to 32 GB), which ensures stable data readout through a single 40 Gbps QSFP+ interface. As of the present moment, notable progress has been achieved, including the successful establishment of 16 data channels between Timepix4 and FPGA that operate error-free and stably at a speed of 5.12 Gbps, which is half of the maximum theoretical speed of 10.24 Gbps. The threshold standard deviation across all pixels is less than 50 e$^{-}$ after equalization. And the clear structural results obtained from X-ray experiments indicate that the  functionality is essentially complete, allowing further testing.}

\notoc
\begin{document}
\maketitle
%change the page number
\clearpage 
\pagenumbering{arabic} 
\setcounter{page}{1}  

\flushbottom

\section{Introduction}
\label{sec:introduction}
Energy-resolved neutron imaging is a non-destructive technique that can be used to investigate the internal structures of materials and analyze stress distributions within crystalline structures \cite{Su_Oikawa_Shinohara_Kai_Horino_Idohara_Misaka_Tomota_2021}. 
The China Spallation Neutron Source (CSNS) detector team has developed an energy-resolved neutron imaging detector based on neutron scintillator screen, precise optical systems, and an optical event camera (TPX3Cam, Amsterdam Scientific Instruments) \cite{yang2021novel,yang2024development,Llopart_Ballabriga_Campbell_Tlustos_Wong_2007}. 
With the increase in proton beam power during the Phase II  of the CSNS, both the peak and total flux of pulsed neutron beams will increase \cite{Fu_2018}.
This necessitates the development of larger imaging areas and higher event-rate readout electronics to meet the demands of the detector upgrade.
\par
Timepix4, a hybrid pixel detector readout ASIC with a sensitive area of \SI{6.94}{cm^2}, offers a tiling capability on four sides and features 55 $\mu\text{m}$ square pixels with sub-nanosecond timestamp binning \cite{llopart2022timepix4}. Additionally, its 160 Gbps readout bandwidth, distributed across 16 data channels with a maximum speed of 10.24 Gbps per channel, supports an event-rate of 2.5 Ghits/s, fully meeting the requirements of the CSNS detector upgrade in terms of position measurement accuracy, time measurement precision, and event-rate.
SPIDR4 is an advanced electronic system developed by NIKHEF for Timepix4 readout, which allows for convenient control and testing of Timepix4 \cite{van2017spidr,Heijhoff_2022}. However, this system needs to be combined with other readout hardware to achieve full-bandwidth data readout, and it is not fully compatible with our existing detector system. To address these limitations, we have developed a dedicated high event-rate electronic system for Timepix4 readout. This paper presents the hardware design, FPGA firmware design, and preliminary testing results of the electronic system.

\section{Hardware design}
\label{sec:hardware}
\begin{figure}[htbp]
\centering
\includegraphics[width=0.65\textwidth]{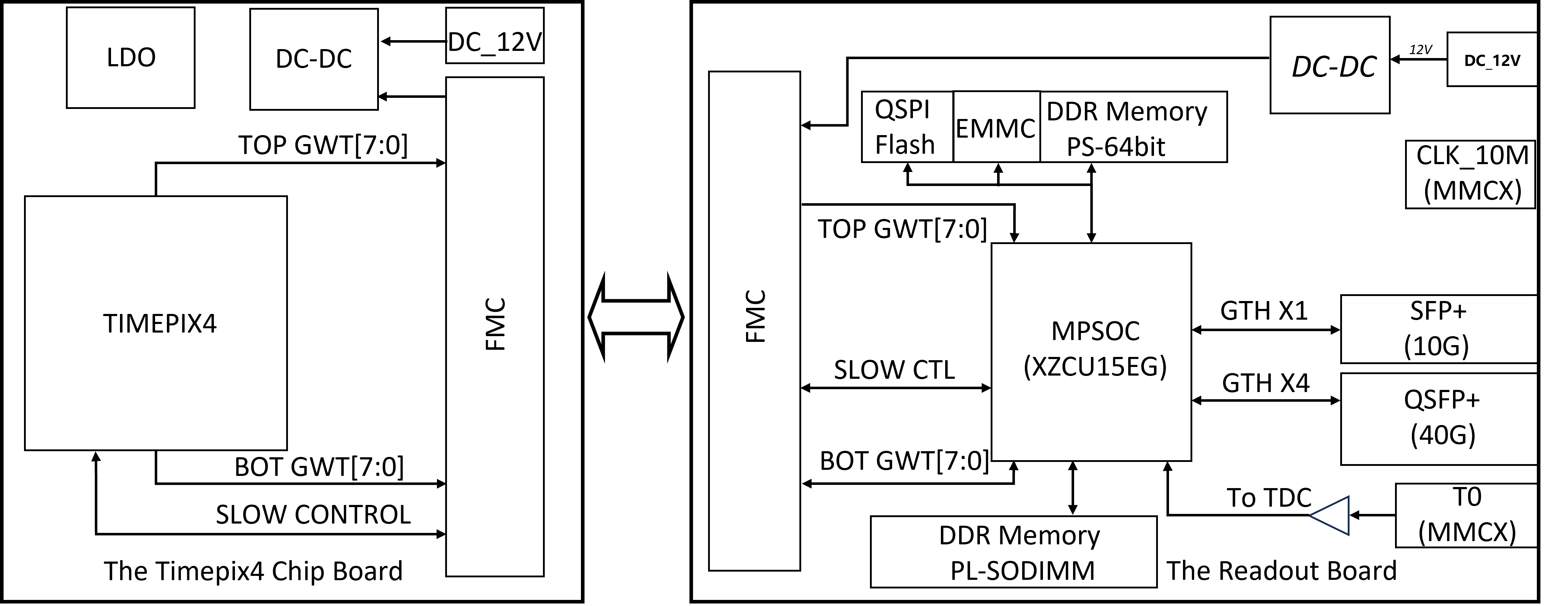}
\caption{Block diagram of the readout electronics.\label{fig:i}}
\end{figure}

In practical applications, the neutron beam does not directly pass through the electronics. Instead, the readout electronics detect only visible light, which is first generated by neutron interactions with the scintillator and subsequently reflected by mirrors. Furthermore, the electronics are shielded with boron-containing materials, reducing the irradiation intensity in their vicinity. Therefore, radiation-hardening designs for our electronics are not required. Figure~\ref{fig:i} illustrates the block diagram of the readout electronics. In light of spatial constraints associated with the installation of the detector, the electronic components have been designed in a compact manner, adhering to dimensions limited to $8\text{ cm} \times 30\text{ cm}$. To facilitate debugging and accommodate future upgrades, the system is partitioned into two parts: the Timepix4 chip board and the digital readout board, which are interconnected through a custom FMC connector. This interface is responsible for transmitting all control signals, high-speed data links, clocks, and power supply.\par

\subsection{Timepix4 chip board}
\label{sec:chipboard}
The Timepix4 chip board is designed to support the Timepix4 ASIC by providing essential power, clock signals, and distributing I/O connections. This board integrates a combination of DC-DC converters and low-dropout regulators (LDOs) to minimize power noise effectively. A dedicated clock chip (SI5345A) \cite{si5345a} provides a programmable reference clock (40 MHz or 320 MHz), synchronized with the readout board. To ensure signal and power integrity, the board uses low-loss PCB material (Megtron 6), minimizes signal vias, and places numerous decoupling capacitors near the power supply pads. Thermal management plays a critical role in ensuring stable operation. The design includes multiple metallized vias that facilitate heat conduction from the Timepix4 chip to the bottom copper layer. Additionally, a thermoelectric cooler (TEC), which exploits the thermoelectric effect for cooling, is integrated beneath the PCB in direct contact with the heat-dissipating copper layer on the opposite side of the Timepix4 chip. This setup is designed to maintain the temperature of the sensor, which is flip-chip bonded to the Timepix4 chip, forming the Timepix4 assembly. Stable temperature control is crucial for minimizing leakage current and noise.\par

\subsection{Digital readout board}
\label{sec:readoutboard}
\begin{figure}[htbp]
\centering
\includegraphics[width=0.8\textwidth]{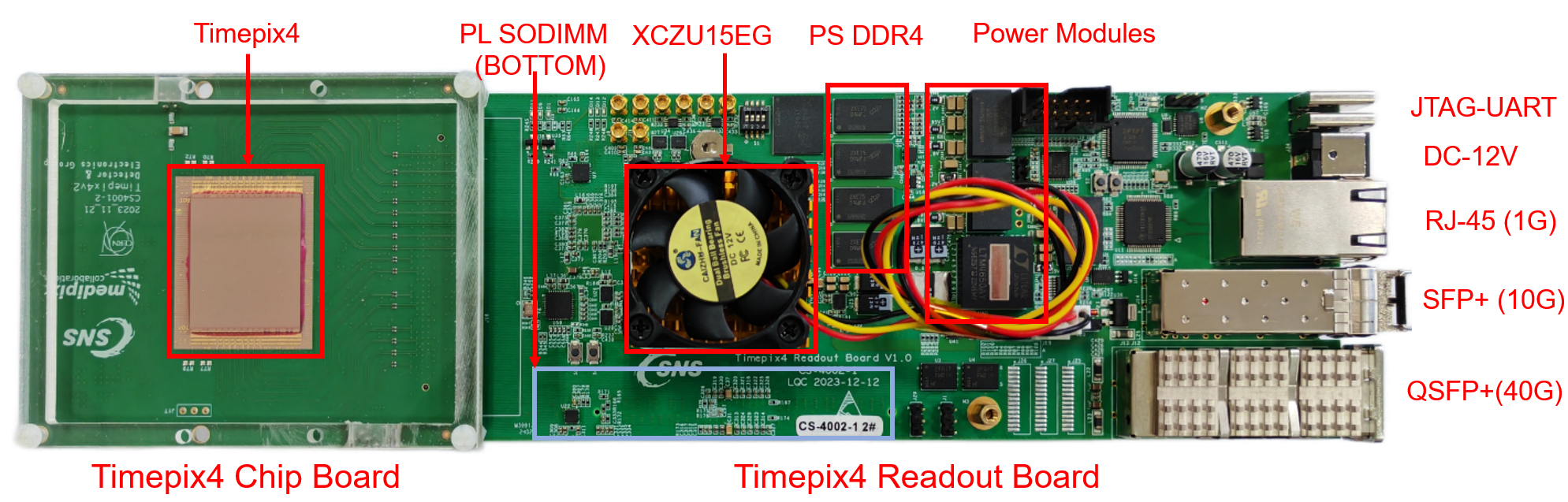}
\caption{The picture of the Timepix4 readout electronics at CSNS.\label{fig:k}}
\end{figure}

The digital readout board controls the Timepix4 chip and processes data from 16 high-speed channels. The core is a ZYNQ-MPSOC (XCZU15EG, Xilinx) \cite{MPSOC} including an ARM-based CPU and FPGA. The CPU operates on a Linux system for  stable control and monitoring of both the Timepix4 chip and the overall system. The chip features 24 pairs of high-speed GTH transceivers, each supporting data rates up to 16.3 Gbps. 16 pairs are dedicated to receiving the data from the Timepix4 chip via the FMC connector, enabling the readout of 16 channels, each capable of up to 10.24 Gbps and compatible with lower speeds such as 5.12 Gbps and 2.56 Gbps. Additionally, the board includes a SODIMM DDR4 memory interface on its underside, capable of providing up to 19.2 GB/s write bandwidth \cite{mpsoc_ddr}, sufficient for handling high-bandwidth experimental data. Figure~\ref{fig:k}  illustrates the Timepix4 readout electronics, comprising the chip board and the digital board.

\section{Firmware design}
\label{sec:firmware}
\begin{figure}[htbp]
\centering
\includegraphics[width=0.6\textwidth]{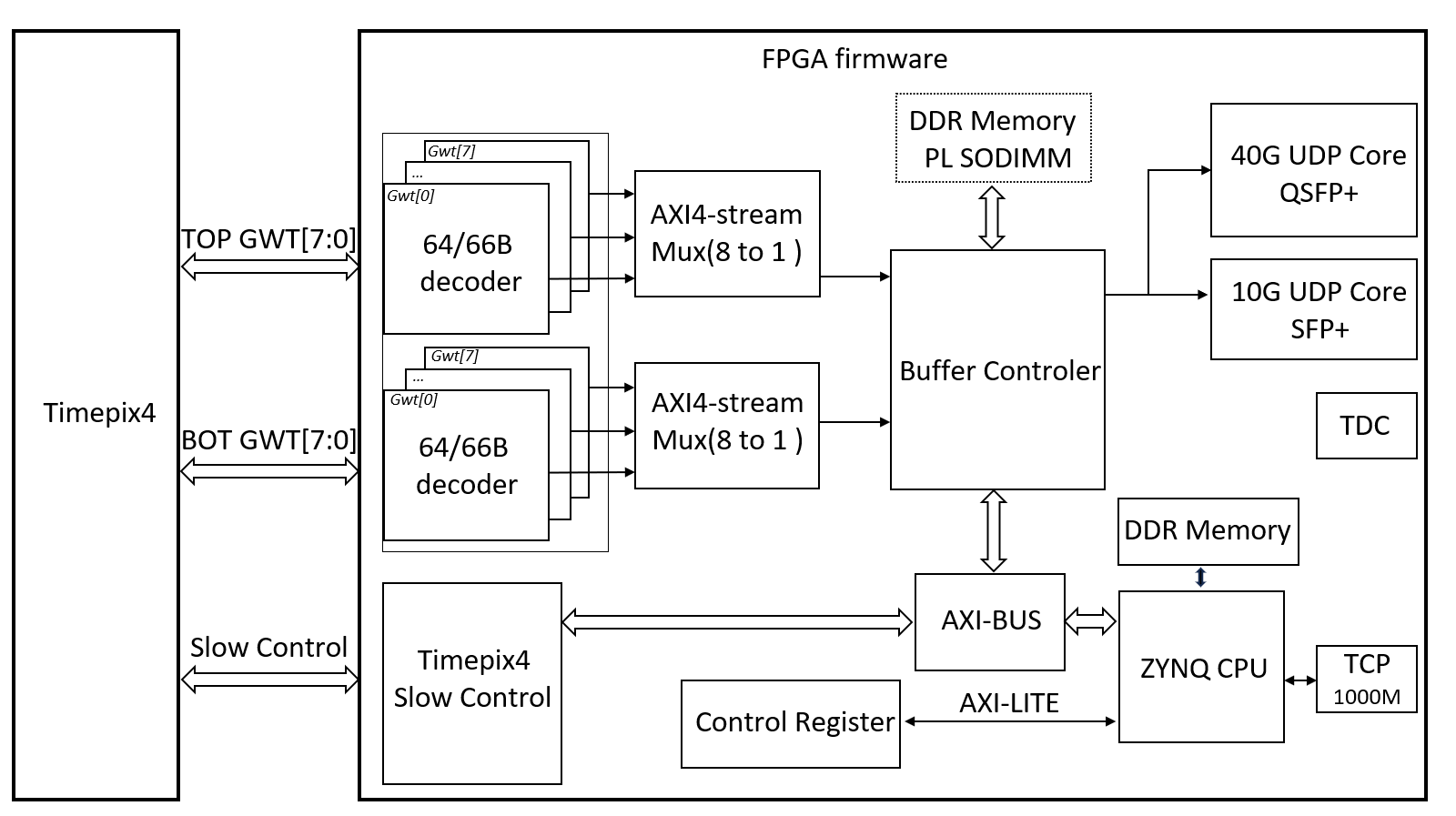}
\caption{Firmware diagram of the Timepix4 readout electronics. (GWT: Gigabit Wireline Transmitter, TOP: Top area of the Timepix4 chip, BOT: Bottom area of the Timepix4 chip). \label{fig:l}}
\end{figure}

Figure~\ref{fig:l} illustrates the architecture and data processing workflow of the firmware system. In this design, experimental data is transmitted via UDP, which is efficient for high-bandwidth transmission but may experience packet loss. To meet the reliability requirements of the control system, control commands and configuration data are transmitted separately using TCP implemented on the PS (Processing System).\par
To facilitate operations such as cross-clock domain processing, multi-channel data merging, arbitration, buffering, and transmission, the data stream is formatted according to the AXI-Stream protocol. Initially, the 16 channel data from both the top and bottom regions are directed into their respective asynchronous FIFOs (First In First Out), ensuring asynchronous processing. Subsequently, this data is merged via two 8-channel AXI-Stream buses managed by the Buffer Control module. This module plays a critical role in monitoring of incoming data bandwidth. In the energy-resolved neutron imaging spectrometer at CSNS \cite{CHEN2024169460}, the pulsed neutron beam is generated at a frequency of 25 Hz, with pulse widths ranging from a few milliseconds to tens of milliseconds and a peak flux of up to \( 10^7 \, \text{n/s/cm}^2 \). When the input bandwidth exceeds the system's maximum real-time readout capacity of 40 Gbps, it temporarily stores excess data in external SODIMM memory. Once the neutron pulse flux peak subsides, buffered data can be efficiently retrieved during idle periods. This multi-tiered caching architecture, leveraging both FIFO and external SODIMM, maximizes the high event-rate and readout bandwidth capabilities of Timepix4.

\section{System testing}
\label{fif:tests}
\subsection{GWT links PRBS-31 tests}
\label{fif:prbs_tests}
\begin{figure}[htbp]
    \centering
    \begin{subfigure}[c]{0.48\textwidth}
        \centering
        \includegraphics[width=\textwidth]{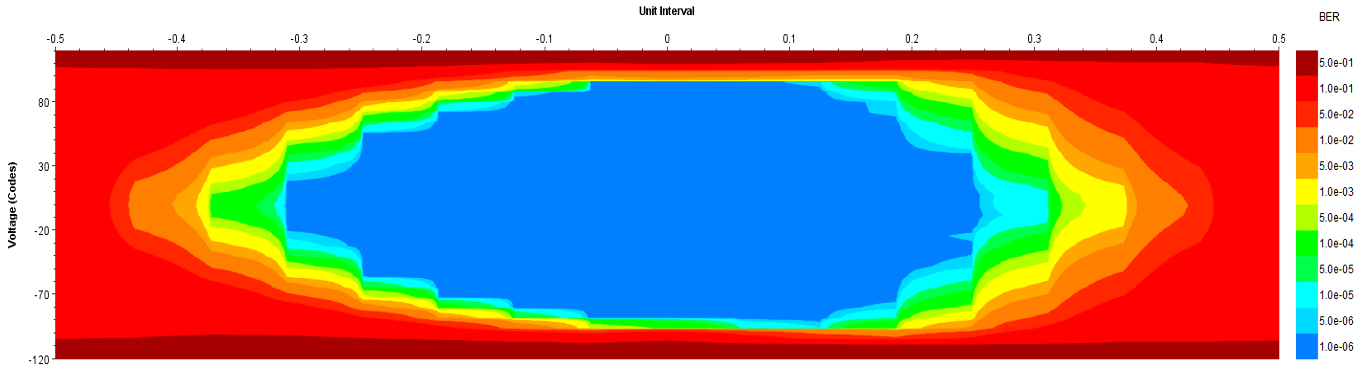}
        \caption{PRBS-31 at 5.12 Gbps: Open UI= 58.82\%}
        \label{fig:sub1}
    \end{subfigure}
    \hfill
    \begin{subfigure}[c]{0.45\textwidth}
        \centering
        \includegraphics[width=\textwidth]{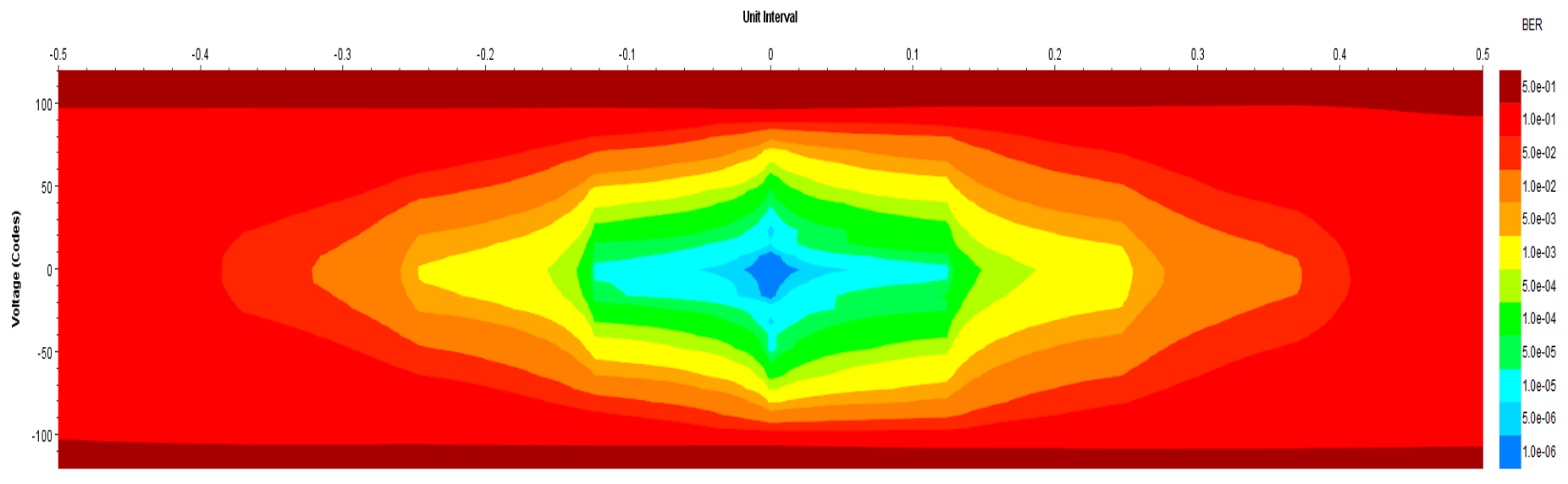}
        \caption{PRBS-31 at 10.24 Gbps: Open UI= 11.11\%}
        \label{fig:sub2}
    \end{subfigure}
    \caption{2D eye scan results for a GWT link using Vivado 2019.2 IBERT default parameters.}
    \label{fig:m}
\end{figure}

% Content for the subsection
The Gigabit Wireline Transmitter (GWT) in Timepix4 is a high-speed serial transmitter circuit designed to convert parallel data from the chip into a single-bit serial data stream. The speed of the GWT links from the Timepix4 chip to the FPGA is a critical parameter in the readout electronics, as it directly influences the overall bandwidth and event-rate of the system. We conducted long duration PRBS-31 tests on 16 GWT channels at a speed of 5.12 Gbps using Vivado IBERT (Integrated Bit Error Ratio Tester) \cite{ibert}, with results indicating no detected errors and a bit error rate (BER) consistently below \(10^{-14}\). Furthermore, we established 16 links operating at 10.24 Gbps, achieving BERs ranging from \(10^{-5}\) to \(10^{-8}\). Figure~\ref{fig:m} presents the 2D eye scan results for one of these GWT links at both speeds: 5.12 Gbps and 10.24 Gbps. At 10.24 Gbps, the eye diagram exhibits a significant reduction in the Unit Interval (UI) opening, accompanied by BER degradation. The issue may be related to parasitic effects from wire-bonding, but its exact cause is still under investigation.

\subsection{Timepix4 equalization}
\label{fif:equalization}
\begin{figure}[htbp]
    \centering
    \begin{subfigure}[c]{.4\textwidth}
        \centering
        \includegraphics[width=\textwidth]{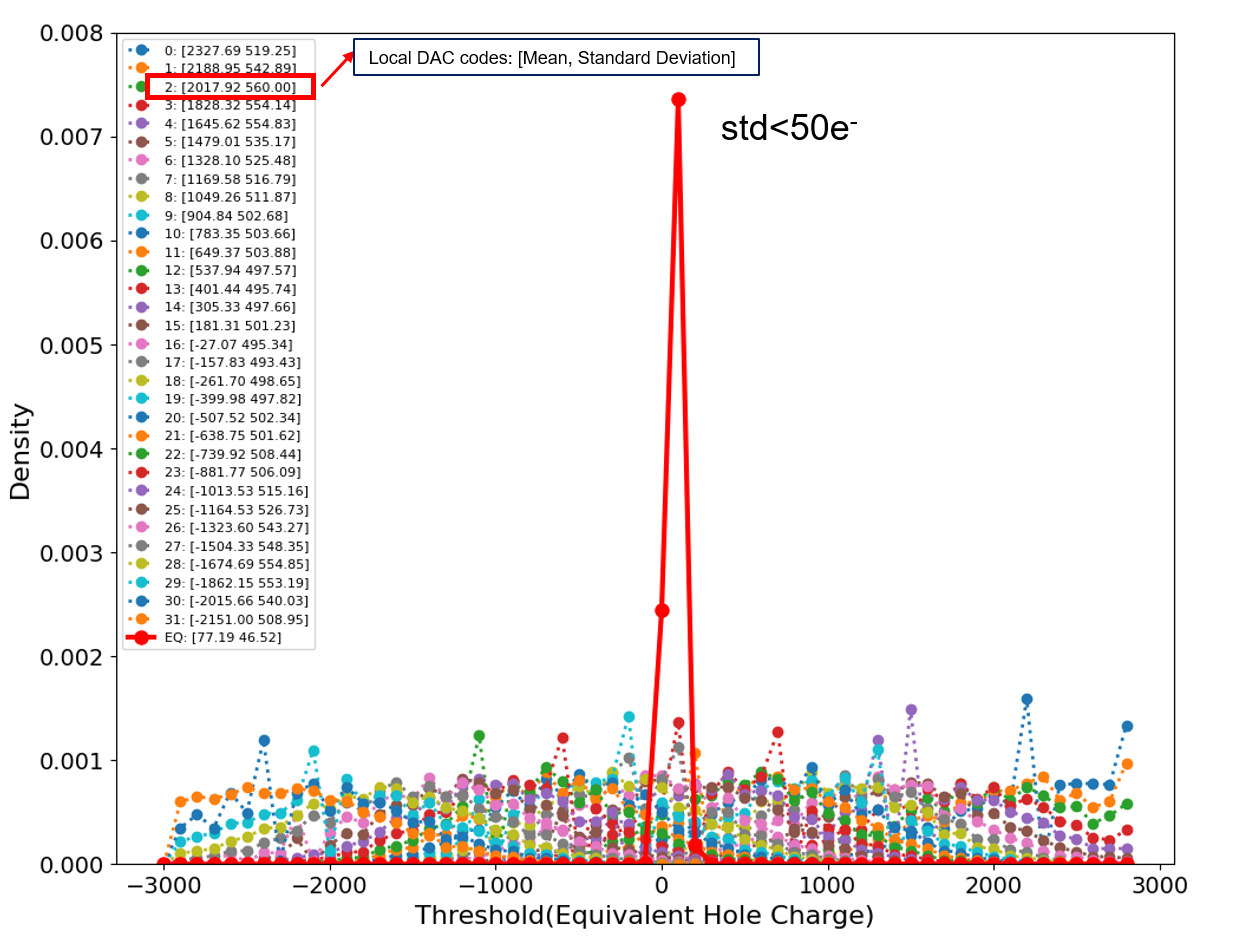}
        \caption{}
        \label{fig:sub3}
    \end{subfigure}
    \qquad
    \begin{subfigure}[c]{0.3\textwidth}
        \centering
        \includegraphics[width=\textwidth]{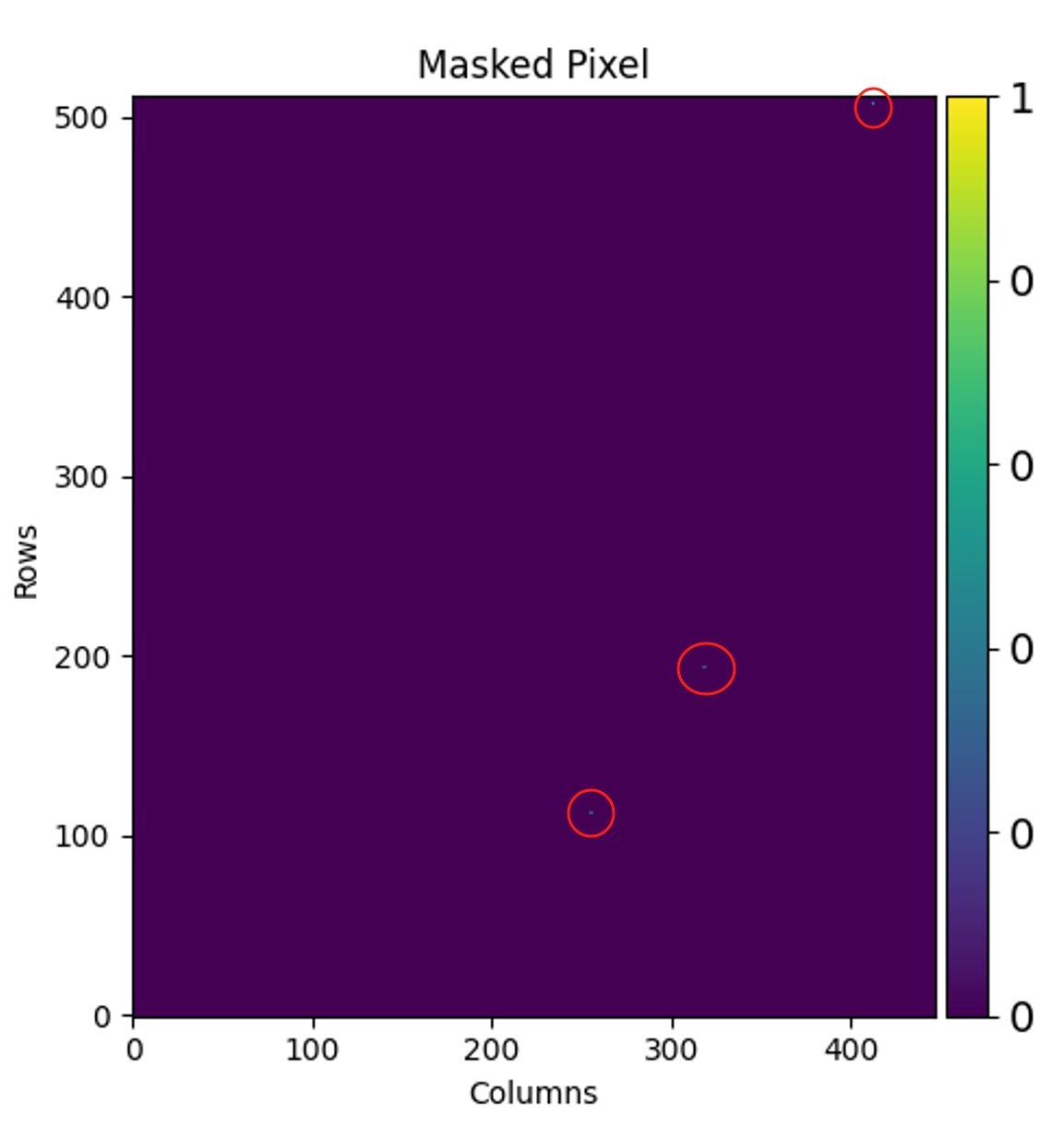}
        \caption{}
        \label{fig:sub4}
    \end{subfigure}
    \caption{(a) Equalization results of a 300 $\mu$m thick p+ in n silicon sensor. The X-axis represents the threshold values (in units of e$^{-}$, electron charge), and the Y-axis denotes the probability density, each point corresponds to a bin with a width of 100 e$^{-}$. (b) Masked pixels (outliers exceeding six times the standard deviation of pixel thresholds).}
    \label{fig:n}
\end{figure}

Due to manufacturing variations, the initial pixel thresholds exhibit observable non-uniformity, with a standard deviation of approximately 500 e$^{-}$. To address this issue, an equalization process is performed. Each pixel in the Timepix4 chip has 32 adjustable local DAC codes for threshold tuning. During the equalization process, each pixel is scanned across its 32 local threshold configurations to identify the optimal DAC codes that minimizes threshold variations. As shown in figure~\ref{fig:sub3}, subplot 0 depicts the pixel threshold distribution when all local thresholds are set to 0, while subplots 1-31 show distributions corresponding to different DAC settings. After equalization, the red subplot demonstrates a significant  reduction in the standard deviation of pixel thresholds to below 50 e$^{-}$, ensuring a more uniform response across all pixels, and only three pixels were masked (figure~\ref{fig:sub4}).

\subsection{X-ray test}
\label{fif:Xrays_test}
\begin{figure}[htbp]
    \centering
    \setlength{\abovecaptionskip}{-5pt} % Reduce space above the captio
    \begin{subfigure}[c]{.3\textwidth}
        \centering
        \includegraphics[width=\textwidth]{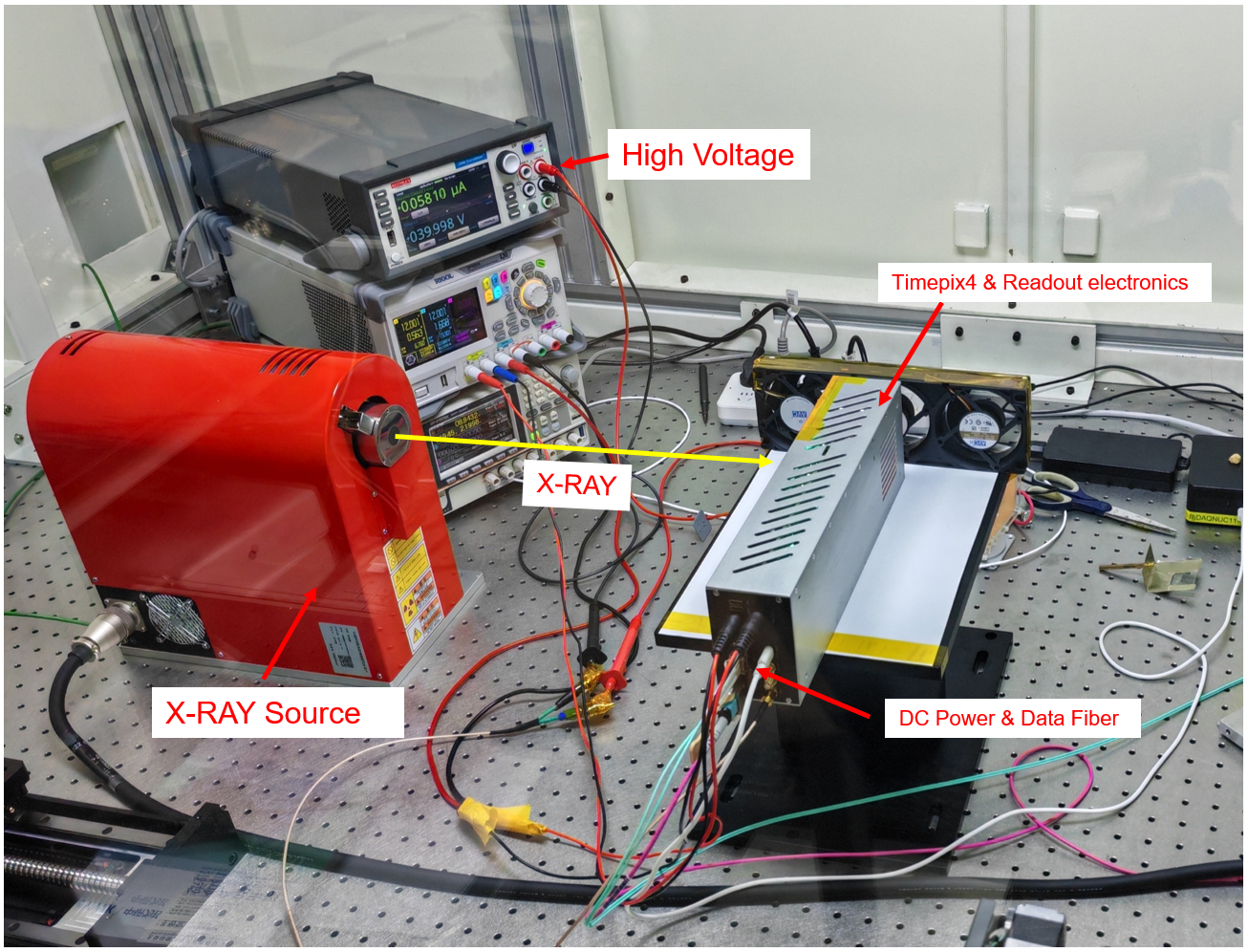}
        \caption{}
        \label{fig:sub7}
    \end{subfigure}
    % \qquad
    \hspace{0.2cm}
    \begin{subfigure}[c]{.25\textwidth}
        \centering
        \includegraphics[width=\textwidth]{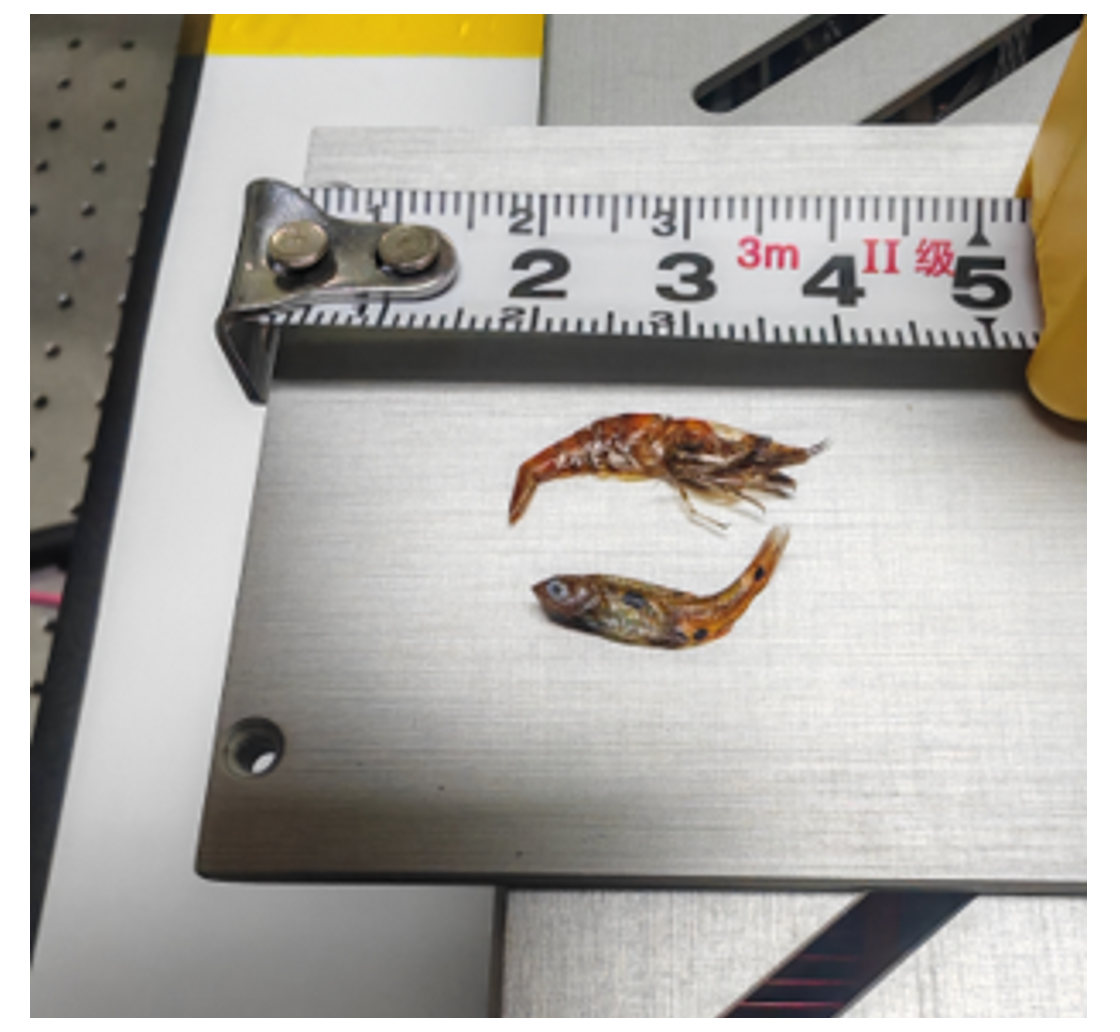}
        \caption{}
        \label{fig:sub5}
    \end{subfigure}
    % \qquad
    \hspace{0.2cm}
    \begin{subfigure}[c]{0.3\textwidth}
        \centering
        \includegraphics[width=\textwidth]{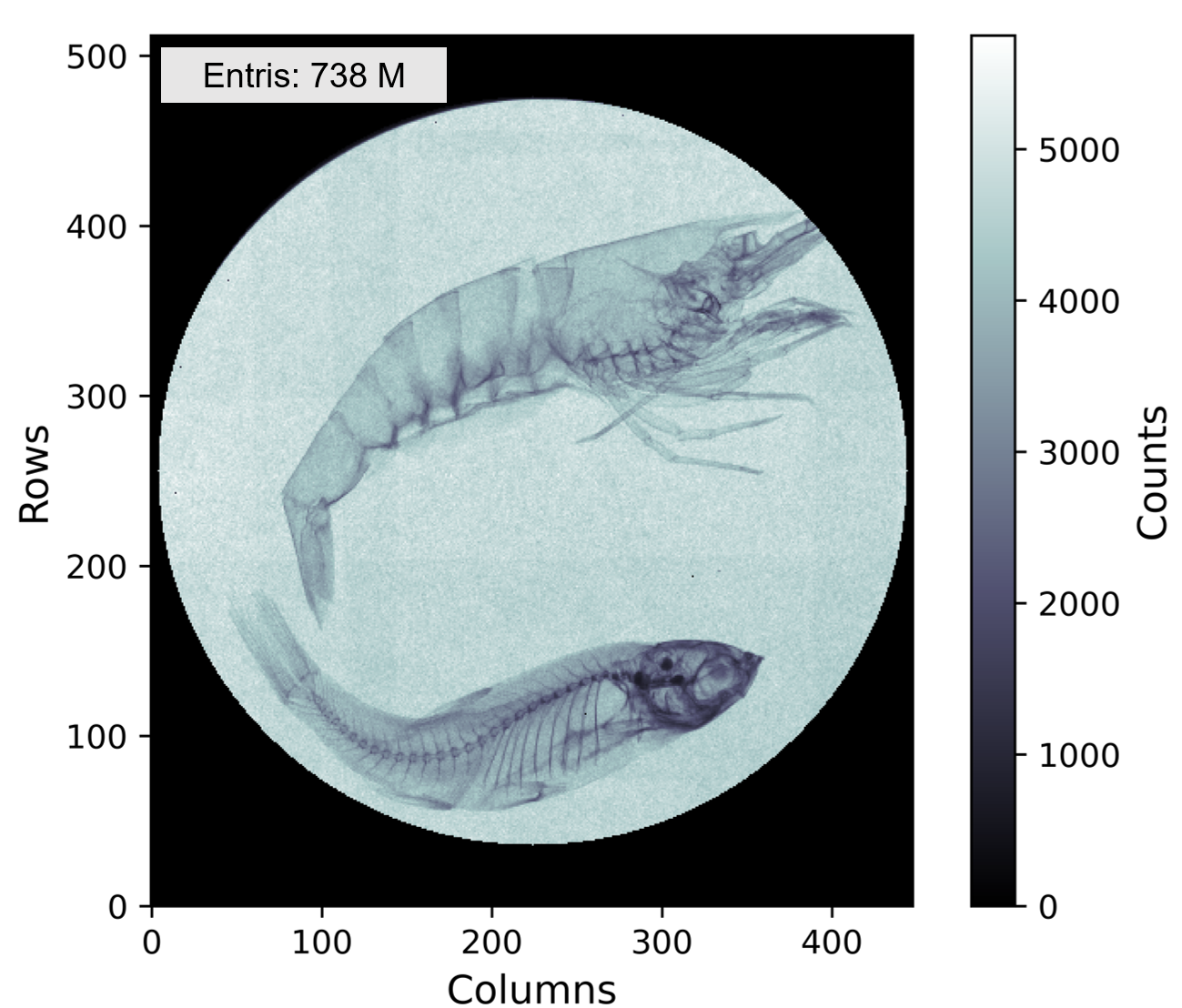}
        \caption{}
        \label{fig:sub6}
    \end{subfigure}
    \caption{(a) Experimental environment, (b) Small fish sample, (c) X-ray imaging of the small fish (raw data). The source was an X-ray tube with tungsten target and a voltage of 100 kV.}
    \label{fig:o}
\end{figure}

 After equalization, we validated the functionality of our entire electronic system through X-ray experiments. Figure~\ref{fig:sub7} illustrates the experimental setup. We used a Hamamatsu (L12161-07) X-ray source \cite{hamamatsu}, positioned at a suitable distance from the sample to approximate a planar source for uniform illumination. A 300 $\mu$m thick p+ in n silicon sensor, flip-chip bonded to the Timepix4 chip and biased at 40 V, was used while the Timepix4 chip operating in event mode. Data was read out using 2 GWT channels at 2.56 Gbps and transmitted via optical fiber to a PC for analysis and storage. We imaged a small fish (figure~\ref{fig:sub5}) and present the results in figure~\ref{fig:sub6}. The figure shows only the X-ray count values acquired over a 60-second exposure, and energy information analysis was not performed in this experiment. In this picture, the fish bone structure can be clearly observed.

\section{Conclusion}
\label{six:conclusion}
We are developing a prototype readout electronics system at CSNS, tailored for high event-rate applications in energy-resolving neutron imaging detectors. At present, the hardware design and firmware development have been completed, along with some basic experimental validations. Through testing of the GWT links, the system currently supports 16 GWT channels at 5.12 Gbps each, providing a total bandwidth of 80 Gbps, which is sufficient for current readout requirements. Equalization results demonstrate that the system achieves a threshold standard deviation of less than 50 e$^{-}$, and the clear structure of the fish's bones obtained by X-ray not only verifies the proper functionality of our readout electronics but also confirms that, after equalization, the electronics exhibit good threshold uniformity. In the future, the electronics are planned to be installed on the energy-resolved neutron imaging detector, and neutron imaging experiments will be conducted.

\appendix
\acknowledgments
This work was supported by the National Natural Science Foundation of China (Grant No. 1222781004), the National Key R\&D Program of China (Grant No. 2024YFE0110001).
The authors would like to thank Dr. Xavi Llopart Cudie from CERN for sharing the Python code for equalization, which was adapted for use in our electronic system.

% Bibtex
\bibliography{biblio.bib}
\end{document}